\documentclass[twocolumn,showpacs,
  superscriptaddress,
  notitlepage,showkeys,
  nofootinbib]{aastex631}
\usepackage{amssymb}
\usepackage{amsmath}
\usepackage{graphicx}
\usepackage{dcolumn}
\usepackage{hyperref}
\usepackage{color,units}
\usepackage{aas_macros}
\usepackage{xspace}
\usepackage[normalem]{ulem} %% for striking out text
\usepackage{soul}
\usepackage{acronym}
\usepackage{tabularx}

\newcommand{\SPA}{School of Physics and Astronomy, Monash University, Vic 3800, Australia}
\newcommand{\OzGravMonash}{OzGrav: The ARC Centre of Excellence for Gravitational Wave Discovery, Clayton VIC 3800, Australia}

\begin{document}
\newacro{GW}{gravitational wave}

\title{Spinning spectral sirens:\\ Robust cosmological measurement using mass-spin correlations in the binary black hole population}

%\title{Spinning spectral sirens: cosmological constraints from multiple features in the binary black holes population}

% \title{Spinning spectral sirens: cosmology with mass-spin correlations in the binary black hole population}

%Spinning spectral sirens: using binary black hole spins to self-calibrate gravitational-wave cosmological inference

%something about a robust/ self-calibrating cosmological measurement by isolating stable features? 

\author{Hui Tong}
\affiliation{\SPA}
\affiliation{\OzGravMonash}
\correspondingauthor{Hui Tong}
\email{hui.tong@monash.edu}

\author{Maya Fishbach}
\affiliation{Canadian Institute for Theoretical Astrophysics, 60 St George St, University of Toronto, Toronto, ON M5S 3H8, Canada}
\affiliation{David A. Dunlap Department of Astronomy and Astrophysics, 50 St George St, University of Toronto, Toronto, ON M5S 3H8, Canada}
\affiliation{Department of Physics, 60 St George St, University of Toronto, Toronto, ON M5S 3H8, Canada}

\author{Eric Thrane}
\affiliation{\SPA}
\affiliation{\OzGravMonash}

\begin{abstract}
Gravitational waves from compact binary mergers provide a direct measurement of luminosity distance, which, in combination with redshift information, serves as a cosmological probe.
In order to statistically infer merger redshifts, the ``spectral standard siren" method relies on features, such as peaks, dips or breaks, in the compact object mass spectrum, which get redshifted in the detector-frame relative to the source-frame. 
However, if the source-frame location of these features evolves over cosmic time, the spectral siren measurement may be biased.
Some features, such as the edges of the pair-instability supernova mass gap, may be more stable than others. 
We point out that binary black hole (BBH) spins, which are not redshifted in the detector-frame, provide a natural way to identify robust mass scales for spectral siren cosmology. 
For example, there is recent evidence for a mass scale in the BBH population that separates slowly spinning from more rapidly spinning BBH mergers, consistent with the lower edge of the pair instability gap. 
Applying our method to data from LIGO-Virgo-KAGRA's third transient catalog, we demonstrate how to isolate this mass scale and produce a robust ``spinning spectral siren" measurement of the Hubble constant: $H_0 = \unit[85^{+99}_{-67}]{km\, s^{-1} Mpc^{-1}}$, or $H_0 = \unit[80^{+60}_{-46}]{km\, s^{-1} Mpc^{-1}}$ when combined with other mass features, such as the $\sim35\,M_\odot$ peak.
We consider the possibility that the source-frame location of the $\sim35\,M_\odot$ peak evolves with redshift, and show that information from black hole spin can be used to mitigate the associated bias for self-calibrating spectral sirens. 
\end{abstract}

%\linenumbers

\section{Introduction}
Starting with the first direct detection of gravitational waves (GWs) in 2015 \citep{gw150914}, the LIGO-Virgo-KAGRA (LVK) Collaboration \citep{aLIGO,aVIRGO,kagra_2013, KAGRA:2013rdx, KAGRA:2020tym} published around 90 compact object mergers by the end of the third observing run \citep{gwtc-3}. More detections from the ongoing fourth observing run are expected to be announced in the near future. 
One exciting application of this expanding GW catalog is standard siren cosmology.
Gravitational waves from compact object mergers provide independent luminosity distance measurements of the sources without any additional distance calibrator. This promises an independent and complementary way to study the expansion history of the Universe \citep{schutz86}, earning these systems the name “standard sirens” in analogy to standard candles \citep{2005Holz}. 

However, due to the degeneracy between the redshift and masses of compact binaries in GW signals, additional efforts are needed to infer the redshifts of GW sources.
Various methods have been proposed. 
Thanks to the detection of the electromagnetic counterpart to the binary neutron star merger GW170817, the source redshift was inferred through its confirmed host galaxy, yielding a measurement of the Hubble constant $H_0=70^{+12}_{-8} \mathrm{ km \ s^{-1} \ Mpc^{-1}}$ \citep{170817}. For sources without an electromagnetic counterpart---but close enough to compared to a complete galaxy catalog---the redshift may be statistically inferred by marginalizing over the potential host galaxies in the GW localization volume
\citep{Del_Pozzo_2012, Chen_2018, Fishbach_2019, Soares_Santos_2019, Gray_2020, gwtc2_cosmo, gwtc3_cosmo, Gray_2022, Gray_2023, Mastrogiovanni_2023, Gair_2023}.
Related techniques, especially useful if the galaxy catalog is incomplete, include comparing the spatial clustering of GW sources and galaxies \citep{Oguri_2016, Bera_2020, Mukherjee_2021, Mukherjee:2022afz}, or leveraging external knowledge of the source merger rate as a function of redshift \citep{Ding_2019, Ye_2021}. For binary neutron stars with tidal signatures in the GW signal, redshifts may be jointly inferred with the neutron star equation of state \citep{Messenger_2012, magnall2024}.

In this paper, we focus on another galaxy-free method for standard siren cosmology known as the ``spectral siren" method, which gains redshift information from the compact binary mass spectrum
\citep{Chernoff_1993, Taylor_2012, Farr_2019, You_2021, Mastrogiovanni_2021, Ezquiaga_2022, mali2024}.
Because the source masses are redshifted in the detector frame, we can simultaneously infer the mass distribution of compact binary mergers alongside the distance distribution and the redshift--distance relation (and hence, cosmological parameters).
The inference is effectively a joint cosmology and population fit using hierarchical Bayesian inference where cosmological parameters can be simultaneously inferred as hyper-parameters. 

Spectral siren analyses are especially promising when there are sharp features in the source-frame mass distribution, such as a narrow Gaussian distribution for neutron star masses~\citep{2012PhRvD..85b3535T} or a steep drop-off in the black hole mass spectrum at the lower edge of the pair-instability mass gap~\citep{Farr_2019}.
Indeed, following the first ten BBH observations, there was preliminary evidence for a cutoff in the mass distribution at component masses $\approx45\,M_\odot$, consistent with the lower edge of the pair-instability supernova (PISN) mass gap~\citep{2017ApJ...851L..25F,2019ApJ...882L..24A}. \cite{Farr_2019} then realized the potential of this feature for cosmology, forecasting a few-percent level measurement of the Hubble expansion at $z \approx 0.8$ with a few years of LIGO-Virgo observations at design sensitivity, and motivating future spectral siren studies.
Theoretically, the PISN gap is a useful feature for spectral siren cosmology not only because of its relatively sharp edges, but because its location in source-frame mass is predicted to evolve only minimally with cosmic time, making it a stable ruler up to high redshifts~\citep{2019ApJ...887...53F}. Furthermore, because PISN are theoretically well-understood, any residual evolution of the source-frame mass gap with redshift could be modeled in order to  calibrate the spectral siren inference. 

However, following the second catalog GWTC-2, the discovery of heavier BBH mergers out to component masses $\approx85\,M_\odot$ \citep{GW190521} washed out preliminary evidence for a sharp cutoff in the BBH mass spectrum, complicating efforts to find the PISN mass gap~\citep{gwtc2_pop}. Either the lower edge of the PISN mass gap starts at much higher masses than previously thought, or the mass gap is contaminated by BHs of non-stellar origin, such as products of previous BBH mergers (so-called second-generation BHs). Although there is no longer a cutoff at $45\,M_\odot$, the mass distribution displays a significant peak at $\approx35\,M_\odot$~\citep{gwtc2_pop}, a feature originally proposed to account for the pulsational PISN pileup preceding the PISN gap~\citep{pp_model}, but possibly due to other aspects of stellar physics~\citep{stevenson}.

This $35\,M_\odot$ peak has driven the spectral siren inference from the latest GW catalogs~\citep{gwtc3_cosmo,mali2024}. 
Nevertheless, the astrophysical origin of this peak remains uncertain, because its location at source-frame masses $\approx35\,M_\odot$ is much lower than expected for pulsational PISN~\citep{2024ApJ...976..121G}. 
While the spectral siren method can still yield cosmological measurements without knowing the astrophysical origin of BBH mass features~\citep{farah2024}, all previous spectral siren analyses have assumed that the source-frame mass distribution does not evolve with redshift, an assumption that is questionable for features like the $35\,M_\odot$ peak whose astrophysical origins are unknown. 
Fundamentally, the problem is that for any individual mass scale, there is an unavoidable degeneracy between its astrophysical evolution with cosmic time and its cosmological redshift~\citep{Ezquiaga_2022,Pierra_2023}.

In this work, we point out that the BBH spin distribution may break this degeneracy between detector-frame masses and redshift, allowing for a robust spectral siren inference.
The BBH spin distribution probably varies with (source-frame) BBH mass \citep{Fishbach:2022lzq, Godfrey:2023oxb,Li_2023yyt,Pierra_2024,antonini2024,afroz2024phasespacebinaryblack}, but unlike masses, spins are not redshifted in the detector frame. 
The BBH spin distribution may therefore be used to identify additional features in the mass distribution that are more reliable than any individual over- or under-density in mass.

In fact, \cite{antonini2024} recently found observational evidence for a mass-spin correlation in which the BBH spin distribution sharply transitions from favoring low spins to favoring spins consistent hierarchical mergers~\citep{Gerosa_2021} at a transition mass of $\approx45\,M_\odot$ in the source frame, assuming standard cosmological parameters inferred from Planck (see \cite{Li_2023yyt,Pierra_2024} for similar findings with different models as well).
They interpret this transition mass as the lower edge of the PISN mass gap, which further motivates us to apply it to spectral siren cosmology. 

We adopt the population models from \citet{antonini2024} to jointly fit the BBH population and cosmology. Moreover, we build on the proposal by \citet{Ezquiaga_2022} to use multiple features in the BBH mass distribution in order to self-calibrate the spectral siren inference. We introduce a method to isolate distinct sources of redshift information from the BBH population by assigning different cosmological parameters to different aspects of the BBH population. This allows us to understand how different aspects of the BBH population contribute to the cosmological inference, isolate the features we trust, and identify sources of systematic uncertainty.

The remainder of this paper is organized as follows. In Sec.~\ref{sec:method}, we introduce the method. In Sec.~\ref{sec:results}, we present the results of analyses on GWTC-3 data and simulated data. Finally, in Sec.~\ref{sec:disucssion}, we present our conclusions and discuss their implications.

\section{Method}\label{sec:method}
\subsection{The spectral siren method}
We perform hierarchical Bayesian inference \citep{2019_Bayesian, Mandel_2019} using \texttt{GWPopulation} \citep{gwpopulation} within the \texttt{Bilby} framework \citep{2019_bilby,2020_bilby}. By Bayes' theorem, the hyper-posterior (of the hyper-parameters $\Lambda$) can be written in terms of the likelihood and prior,
\begin{equation}
p(\Lambda|\{d\})=\frac{\mathcal{L}(\{d\}|\Lambda)\pi(\Lambda)}{\int\mathcal{L}(\{d\}|\Lambda)\pi(\Lambda)\text{d}\Lambda}.
\end{equation}
In a spectral siren analysis, the hyper-parameters $\Lambda=\{\Lambda_\text{pop}, \Lambda_{c}\}$ consist of population parameters $\Lambda_\text{pop}$, which describe the astrophysical mass, spin and redshift distributions, and cosmological parameters $\Lambda_{c}$.
The likelihood for the GW data in the case of N independent detections, $\{d_i\}$, given the assumed hyper-parameters, marginalizing over the merger rate, is
\begin{equation}
\mathcal{L}(\{d\} | \Lambda)=\prod_{i}^{N}\frac{\int\mathcal{L}(d_i|\theta,\Lambda_c)\pi_\text{pop}(\theta|\Lambda_\text{pop})\text{d}\theta}{\alpha(\Lambda)},
\end{equation}
where $\theta$ are the event-level parameters (e.g., BBH masses), $\mathcal{L}(d_i|\theta,\Lambda_c)$ is the individual event likelihood and $\pi_\text{pop}$ is the population model.
The normalization factor accounts for observational selection effects,\footnote{The curious $\text{d}d$ term in Eq.~\ref{eq:alpha} is part of an integral over all possible data.}
\begin{equation}\label{eq:alpha}
\begin{aligned}
\alpha(\Lambda)=&\int_{d>\mathrm{threshold}}\text{d}d\int\text{d}\theta\mathcal{L}(d|\theta,\Lambda_c)\pi_\text{pop}(\theta|\Lambda_\text{pop})\\
=&\int\text{d}\theta \, p_{det}(\theta|\Lambda_c)\pi_\text{pop}(\theta|\Lambda_\text{pop}).
\end{aligned}
\end{equation}
Here, $p_\text{det}(\theta|\Lambda_c)$ is the probability of detecting a BBH merger with parameters $\theta$ in a universe with cosmological parameters $\Lambda_c$.
Previous spectral siren analyses usually modeled the population distribution $\pi_\text{pop}$ in terms of source-frame masses and redshifts,  including these properties in $\theta$. In this case, the likelihood of a single GW event $\mathcal{L}(d_i|\theta,\Lambda_c)$ depends also on the corresponding cosmological parameters $\Lambda_c$.

The single-event likelihood $\mathcal{L}(d_i|\theta)$ does not depend on cosmological assumptions if we choose $\theta$ to represent only the detector-frame GW parameters, like redshifted masses $m_\mathrm{det}$ and luminosity distance $d_L$.
If our population model $\pi_\text{pop}$ described the detector-frame population with sufficient flexibility, we would not learn anything about cosmology. 
On the other hand, for a given cosmological model, we can transform any source-frame population model into the implied detector-frame distribution, coupling the cosmological parameters to the inferred detector-frame distribution.

%We can consider an in-between scenario in which $\theta$ consists of a mixture of source-frame and detector-frame parameters. For example, we can model the distribution of source-frame masses and luminosity distances (rather than redshifts). The resultant cosmological inference would leverage features of the mass spectrum. On the other hand, a population model of source-frame mass and redshift will also include cosmological information from features in the redshift distribution.
Explicitly factoring the population model $\pi_\text{pop}$ in terms of detector-frame masses $m_\mathrm{det}$, luminosity distances $d_L$ and spins (parameterized with the effective inspiral spin parameter $\chi_\mathrm{eff}$), we see that multiple GW source parameters carry information about cosmology:
\begin{equation}\label{eq: original cosom likelihood}
\begin{aligned}
&\mathcal{L}(d_i|\Lambda)=\frac{1}{\alpha(\Lambda)}\int \text{d}\theta \mathcal{L}(d_i|m_{\text{det}},d_L,\chi_{\rm{eff}})\pi(d_L|\Lambda_\text{pop}, \Lambda_{c, z})\\
&\pi(m_{\text{det}}|\Lambda_\text{pop}, \Lambda_{c,m}, d_L)\pi(\chi_{\rm{eff}}|\Lambda_\text{pop}, \Lambda_{c,s}, m_{\text{det}}).
\end{aligned}
\end{equation}
Indeed, there are three contributions to the cosmological inference in our population model, which we isolate by allowing them each to have a distinct set of cosmological parameters ($\Lambda_{c,z}$, $\Lambda_{c, m}$ and $\Lambda_{c,s}$):
\begin{enumerate}
    \item \textit{Luminosity distance}---\textit{redshift}. Given a redshift distribution model with population hyper-parameters $\pi(z|\Lambda_{pop,z})$, the corresponding luminosity distance model $\pi(d_L|\Lambda_{pop,z}, \Lambda_{c,z})$ can be calculated by a change of variables. Constraints on the inferred cosmological parameters $\Lambda_{c,z}$ come from features in the modeled redshift distribution and their corresponding signatures in the observed luminosity distance distribution~\citep{Ding_2019,Ye_2021}.
    
    \item \textit{Detector-frame mass}---\textit{source-frame mass}. Given a source-frame mass model with population hyper-parameters $\pi(m_{\text{source}}|\Lambda_{pop,m})$, the corresponding detector-frame mass distribution at a given luminosity distance $\pi(m_{\text{detector}}|\Lambda_{pop,m}, \Lambda_{c,m}, d_L)$ can be calculated given cosmological hyper-parameters $\Lambda_{c,m}$. The inference of $\Lambda_{c,m}$ relies on features in the mass model~\citep{2012PhRvD..85b3535T,Farr_2019}.
    
    \item \textit{Effective spin}. Although BBH spin parameters are not redshifted in the detector frame, the population-level correlation between source-frame masses and spins means that the conditional spin distribution depends on cosmological parameters as well. Presuming the existence of the transition mass $m_t$ \citep{antonini2024} in the source frame, there is a redshifted, detector-frame transition mass that depends on cosmological parameters $\Lambda_{pop,s}$, so we write $\pi(\chi_{\rm{eff}}|\Lambda_{pop,s}, \Lambda_{c,s}, m_{\text{det}}, d_L)$.
\end{enumerate}

Physically, there is only one set of cosmological parameters, and if we knew our population models were adequately specified, we could assume 
\begin{align}
    \Lambda_{c,m} = \Lambda_{c,s} = \Lambda_{c,z} .
\end{align}
By independently assigning different cosmological hyper-parameters $\Lambda_{c,z}$, $\Lambda_{c, m}$ and $\Lambda_{c, s}$, we can learn how different features in the compact binary population contribute to the cosmological inference, helping to hedge against model misspecification.\footnote{For a review of model misspecification in gravitational-wave astronomy, see \cite{wmf}.}
There are inevitable correlations between these different sets of cosmological hyper-parameters. For example, the detector-frame mass model is conditioned on $\Lambda_{c,m}$ and luminosity distance while the luminosity distance distribution depends on $\Lambda_{c,z}$.
However, the correlations are small if the models are flexible enough to fit the data, allowing us to effectively isolate the cosmological constraints coming from the different features.

\subsection{Models}
Following \cite{gwtc-3_pop,gwtc3_cosmo}, we adopt the \textsc{Power Law + Peak} model \citep{pp_model}, referred to here as \textsc{Power Law + fixed Peak}, as the fiducial model for the distribution of black-hole masses. The source-frame mass distribution is assumed to be of the form $p(m_1, m_2)\propto p(m_1)p(m_2 | m_1)$ with $p(m_1)$ a mixture of a power law component and a Gaussian peak,
\begin{equation}
\begin{aligned}
p(m_1|m_{\rm{min}},&m_{\rm{max}},\alpha,\lambda_g,\xi_z,\mu_g,\sigma_g,z)=\\
(1&-\lambda_g)\mathcal{P}(m_1|m_{\rm{min}},m_{\rm{max}},-\alpha)+\\
&\lambda_g \mathcal{G}(m_1|\mu=\mu_g,\sigma=\sigma_g).
\end{aligned}
\end{equation}
Here, $\mathcal{P}(m_1|m_{\rm{min}},m_{\rm{max}},-\alpha)$ is a truncated power law described by slope $-\alpha$ and lower and upper bounds $m_{\rm{min}},m_{\rm{max}}$ and $\mathcal{G}$ is a Gaussian distribution. The secondary mass follows another truncated power law,
\begin{equation}
    \pi(m_2|m_1,m_{\rm{min}},\beta)=\mathcal{P}(m_2|m_{\rm{min}},m_1,\beta).
\end{equation}
A smoothing factor, described by $\delta_m$, is applied to the lower end of both $m_1$ and $m_2$ distributions, the explicit expression for which can be found in Eq. (B6) and Eq. (B7) of \cite{gwtc2_pop}.
A major concern for spectral siren cosmology is that the features in the source-frame mass distribution may evolve with redshift \citep{Ezquiaga_2022, Pierra_2023}.
Thus, we also explore a simple variation of the \textsc{Power Law + Peak} model in which the center of the Gaussian peak may linearly evolve with redshift. We refer to this model as the \textsc{Power Law + evolving Peak} model,
\begin{equation}
\begin{aligned}
p(m_1&|m_{\rm{min}},m_{\rm{max}},\alpha,\lambda_g,\xi_z,\mu_g,\sigma_g,z)=\\
(1&-\lambda_g)\mathcal{P}(m_1|\Lambda)+\lambda_g \mathcal{G}(m_1|\mu=(1+\xi_z z)\mu_g,\sigma=\sigma_g).
\end{aligned}
\end{equation}
Here $\Lambda$ represents the hyper-parameters needed for the power law component, $\mu_{g}$ is the center of the Gaussian peak at $z = 0$, $\xi_z$ is the slope of its redshift evolution, and $\sigma_g$ is its width. Note that $z$ is the source redshift, not a population hyper-parameter.

For the redshift distribution, we parameterize the merger rate as a function of redshift following \cite{Madau_2014},
\begin{equation}
    \psi(z|\gamma,\kappa,z_p)=[1+(1+z_p)^{-\gamma-\kappa}]\frac{(1+z)^\gamma}{1+[(1+z)/(1+z_p)]^{\gamma+\kappa}}.
\end{equation}
The corresponding redshift distribution of BBHs (per unit redshift interval) is then
\begin{equation}\label{model:redshift}
    p(z|\gamma,\kappa,z_p)\propto\frac{1}{1+z}\frac{dV_c}{dz}\psi(z|\gamma,\kappa,z_p).
\end{equation}
We use the mass-dependent spin model introduced in \cite{antonini2024}. We model the effective spin distribution conditioned on the primary mass of BBHs,
\begin{equation}\label{model:spin}
\begin{aligned}
& \pi(\chi_\mathrm{eff}|m_1,\Lambda) \\
& =
\begin{cases}
\mathcal{N}(\chi_\mathrm{eff}|\mu,\sigma) & (m_1<m_t) \\
\zeta\,\mathcal{U}(\chi_\mathrm{eff}|w) + (1-\zeta) \mathcal{N}_{\rm u}(\chi_\mathrm{eff}|\mu_{\rm u}, \sigma_{\rm u}) & (m_1 \geq m_t).
\end{cases}
\end{aligned}
\end{equation}

For the cosmological model, we consider a flat $\Lambda$CDM Universe with $\Omega_m$ fixed at 0.3065 \citep{Planck2015}. We only consider $H_0$ as a free parameter, given that current low-redshift data are not yet sensitive to other cosmological parameters.

\section{Results}\label{sec:results}
\subsection{GWTC-3 results}\label{sec:GWTC3-results}
Following \cite{gwtc-3_pop}, we include 69 BBHs in our analyses with the threshold\footnote{Note the event list is different from \cite{gwtc3_cosmo} since we use a different selection criterion (based on FAR instead of signal-to-noise ratio). We model selection effects during hierarchical Bayesian inference. See Appendix~\ref{appendix:alternative dataset} for a comparison between different choices of selection criteria.} of false alarm rate $\text{FAR}\textless\unit[1]{yr^{-1}}$.
For each event, we use parameter estimation samples from the GWTC-3 data release. We use the combined set of injections for GWTC-3 search sensitivity estimates to take the selection effect into account \citep{GWTC-3_release}. We include other details of our analysis, including a full description of priors, in Appendix.~\ref{appendix:priors}.

\begin{figure*}
    \centering
    \includegraphics[width=1.0\linewidth]{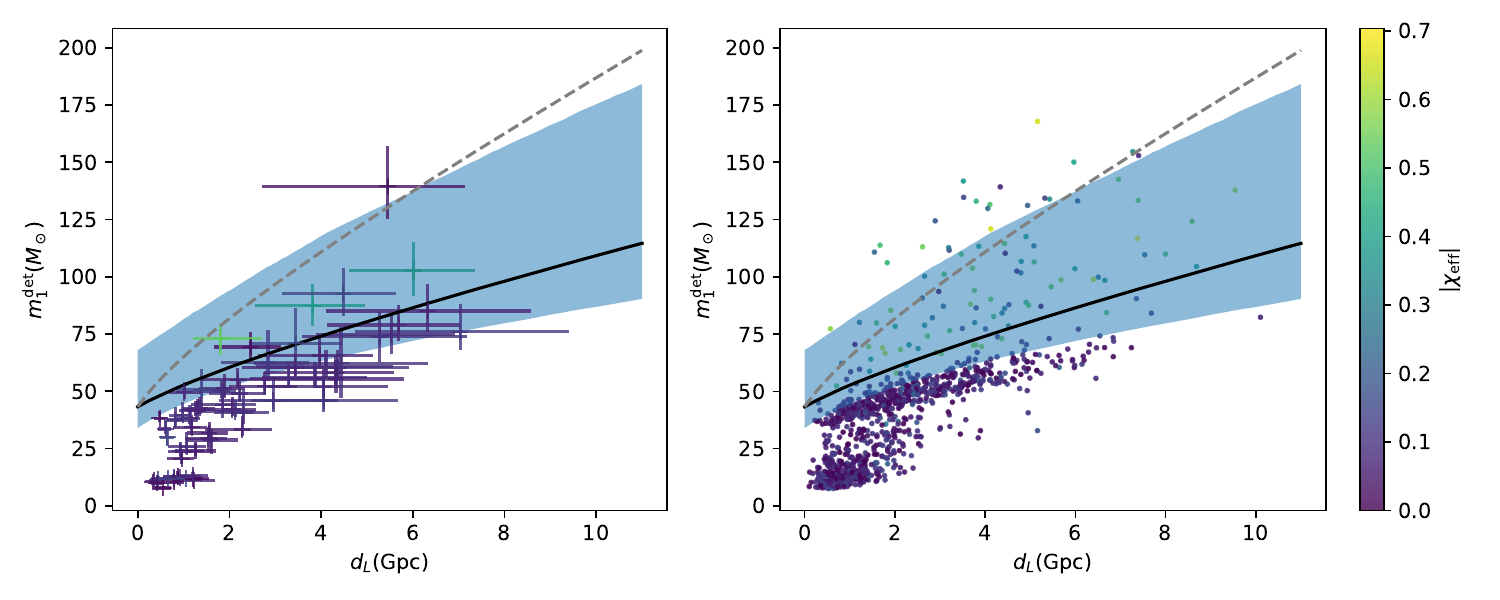}
    \caption{Primary masses and luminosity distances colored according to $\chi_{\mathrm{eff}}$.
    The left-hand plot shows results from GTWC-3 assuming a population-informed prior (error bar shows 90\% credible interval).
    The right-hand plot is for a simulated population of BBH mergers detected by O3b-like detection sensitivity. The black curve shows the redshifted transition mass $m_t^\text{det}$.
    According to the \cite{antonini2024} model, BBH events with large $\chi_\text{eff}$ (yellow-green) should fall above this transition.
    The shaded blue area shows the median and 90\% credible interval for $m_t^\text{det}$.
    When $d_L \approx 0 $, the uncertainty is dominated by the population analysis, which is carried out assuming the \textsc{Power Law + fixed Peak} and mass-dependent spin model.
    At larger values of $d_L$, the uncertainty is dominated by uncertainty in $H_0$; see Fig.~\ref{fig:69_H0_comparison}). The black curve can be contrasted with the dashed gray curve where we assume a source-frame transition mass equal to $43M_\odot$ and an improbable value of the Hubble constant $H_0=200\mathrm{ km \ s^{-1} \ Mpc^{-1}}$. 
    While the high-$\chi_\text{eff}$ events consistently fall above the black curve (with a plausible $H_0$ value), many are below the dashed gray curve with an implausible value.
    The simulated population assumes no mass evolution with all other parameters the same as those true values listed in Appendix~\ref{appendix:injection model}.}
    \label{fig:Demostration mass spin correlation}
\end{figure*}

Figure~\ref{fig:Demostration mass spin correlation} shows scatter plots of detector-frame primary mass $m_1^\text{det}$ versus luminosity distance $d_L$ for events in GWTC-3 (left) and for simulated data (right).
Each event is color-coded according to its $\chi_\text{eff}$ value.
The black curve shows the expected transition mass in the detector frame, given $H_0$
\begin{align}
    m_t^\text{det}(d_L | H_0) = 
    m_t \big(1 + z(d_L | H_0)\big) .
\end{align}
According to the hypothesis put forward in \cite{antonini2024}, events above this curve are more likely to exhibit large values of $\chi_\text{eff}$ (greenish yellow) compared to events below it.
The blue shaded region represents the 90\% credible uncertainty for $m_t^\text{det}$.
For $d_L \approx 0$, this uncertainty comes from the population estimate of $m_t$, but when $d_L$ is large, the uncertainty coming from our imperfect knowledge of $H_0$.

While the high-$\chi_\text{eff}$ GWTC-3 events consistently fall above the black curve, they are all below the dashed gray curve, which shows how $m_t^\text{det}$ would evolve with a Hubble constant $H_0=200\mathrm{ km \ s^{-1} \ Mpc^{-1}}$ inconsistent with the GW data.
The plot highlights how spin measurements---which are not affected by redshift---can be used to help constrain $H_0$, even if it turns out that $\pi(m_1)$ evolves with redshift.

\begin{figure}
    \centering
    \includegraphics[width=1\linewidth]{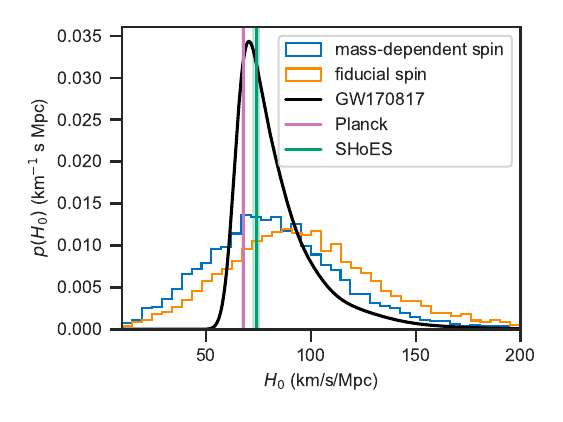}
    \caption{Comparison of $H_0$ using the mass-dependent spin model (blue) versus the fiducial spin model (orange). The mass model for both analyses is \textsc{Power Law + fixed Peak}, and the redshift model is given by Eq.~\ref{model:redshift}. For reference, we also plot the constraints on $H_0$ from the bright standard siren GW170817 \citep[black,][]{170817, GW170817_cosmo} as well as the $H_0$ measurements from the CMB by Planck \citep[pink band,][]{Planck2015} and from the local distance ladder by SH0ES \citep[green band,][]{Riess_2019}, with shaded areas identifying the 68\% credible interval.}
    \label{fig:69_H0_comparison}
\end{figure}

In Fig.~\ref{fig:69_H0_comparison}, we show the $H_0$ posterior under the fiducial \textsc{Power Law + fixed Peak} model (orange). We assume the mass and redshift components of the population model share the same $H_0$ ($H_{0,m} = H_{0,z}$). A uniform prior on $H_0$ is adopted in the range of [10, 200] $\mathrm{ km \ s^{-1} \ Mpc^{-1}}$. In blue, we simultaneously fit the BBH spin distribution using the mass-dependent spin model introduced in Eq.~\ref{model:spin}, once again assuming a single $H_0$ for all population features, $H_{0,s} = H_{0,m} = H_{0,z}$. For the posterior in orange, the spin distribution is assumed to be the same as the fiducial spin priors used by the LVK for single event parameter estimation (referred to as the fiducial spin model hereafter). The fiducial spin priors are independent of any cosmology assumptions, and do not significantly impact the cosmological inference. In contrast, the spin distribution in the mass-dependent spin model transitions from a single Gaussian distribution to a mixture of a uniform and and a Gaussian after a certain (source-frame) mass scale $m_t$, which we simultaneously infer. Because this mass scale is redshifted in the detector frame, it provides an additional source of cosmological information.

\begin{figure}
    \centering
    \includegraphics[width=1\linewidth]{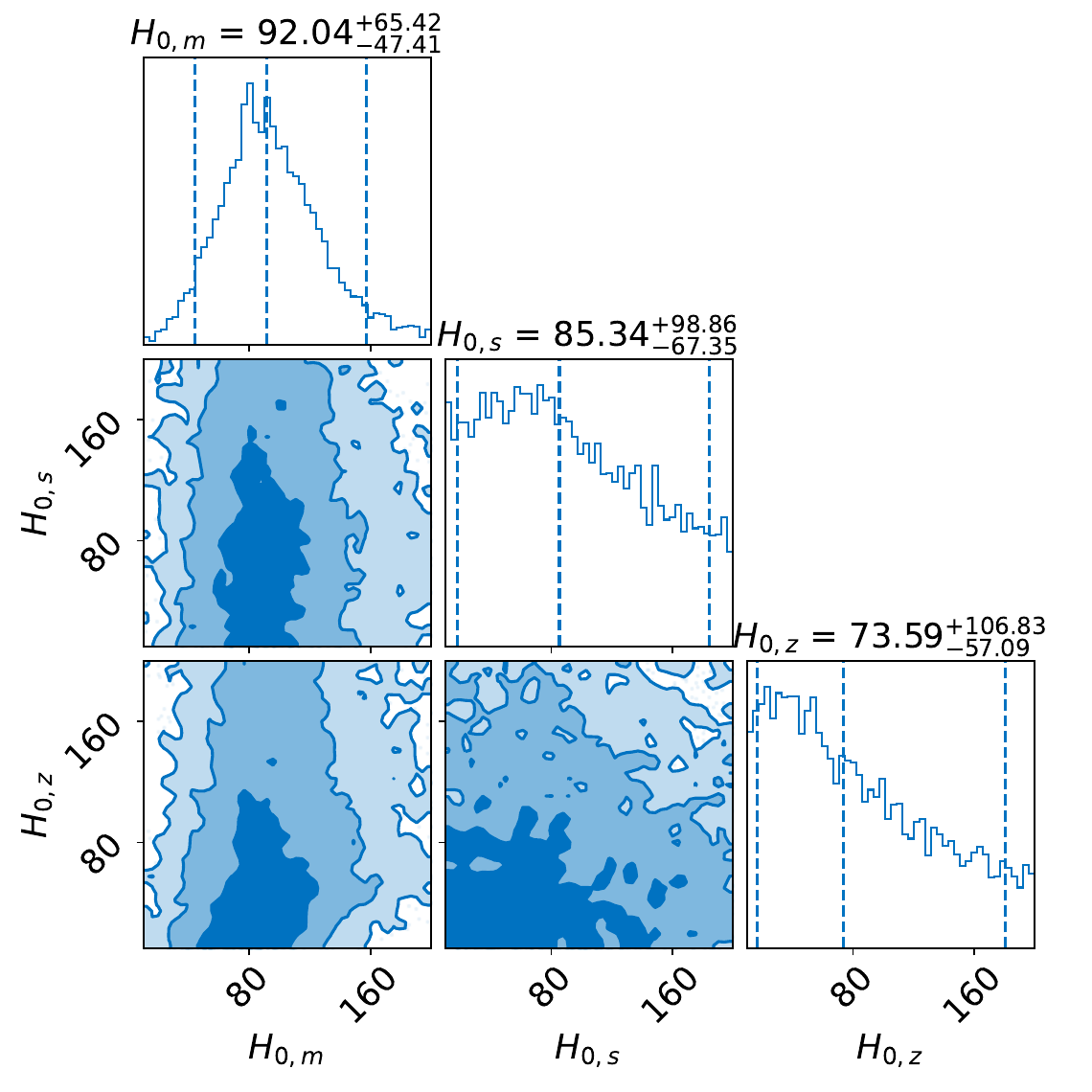}
    \caption{Constraints on $H_0$ from different parts of the population model using the \textsc{Power Law + fixed Peak} mass model and mass-dependent spin model. The Hubble parameter $H_0$ is measured in units of $\unit[]{km\,s^{-1}Mpc^{-1}}$.}
    \label{fig:69_H0_multi}
\end{figure}

Including the extra cosmological information from the spin-mass correlation improves the $H_0$ measurement slightly from $H_0=93^{+67}_{-52}$ under the fiducial model to $H_0=80^{+60}_{-46}$ (all measurements are median and 90\% credible interval).
However, the main advantage of introducing an additional mass scale $m_t$ from the spin-mass correlation is the added robustness compared to the fiducial mass scales in the population (e.g., the Gaussian peak whose physical origin is unknown).

The addition of $m_t$ allows us to cross-check our fiducial $H_0$ results. We isolate the constraints on $H_0$ coming from different features in the population, using the method introduced in Sec.~\ref{sec:method}. 
In Fig.~\ref{fig:69_H0_multi}, we show the result obtained by assigning different $H_0$ to the different population features: the \textsc{Power Law + fixed Peak} mass model, the redshift model, and the mass-dependent spin model. For each estimate of $H_{0}$, we use independent uniform priors on the interval $[10, 200]\ \mathrm{ km \ s^{-1} \ Mpc^{-1}}$. 
The different $H_0$ posteriors coming from different parts of the population are consistent with each other within their broad uncertainties. 
As expected, the primary mass distribution provides the best constraints on $H_0=92^{+65}_{-47}\ \mathrm{ km \ s^{-1} \ Mpc^{-1}}$.
However, the spin-mass correlation and the redshift distribution also contribute some information. Using the spin-mass correlation as a cosmological ruler yields $H_0=85^{+99}_{-67}\ \mathrm{ km \ s^{-1} \ Mpc^{-1}}$, which is encouragingly consistent with the value obtained from the primary mass distribution.

Next, we consider a mass distribution that evolves with redshift. This analysis motivates our simulation in the following subsection. If the transition mass for the spin-mass correlation is indeed the lower edge of the PISN mass gap, we expect its redshift evolution to be minimal. However, we have little prior constraints on how the features in the primary mass distribution, such as the location of the Gaussian peak, may evolve with redshift, so we consider the \textsc{Power Law + evolving Peak} model.
In the \textsc{Power Law + evolving Peak} model, the evolution of the Gaussian peak is determined by $\xi_z$.
We assume a uniform prior on $\xi_z$ between $[-1,1]$.

\begin{figure}
    \centering
    \includegraphics[width=1\linewidth]{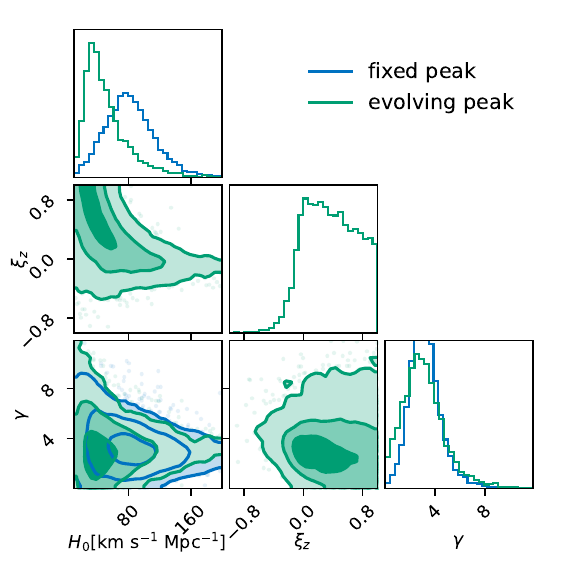}
    \caption{Posteriors of $H_0$, the mass evolution parameter $\xi_z$, and the low-redshift merger rate evolution parameter $\gamma$ inferred under the \textsc{Power Law + fixed Peak} model (blue) and the \textsc{Power Law + evolving Peak} model (green). Both analyses assume the mass-dependent spin model.}
    \label{fig:nomulti_evolve_corner}
\end{figure}

Fig.~\ref{fig:nomulti_evolve_corner} compares the $H_0$ posterior inferred by the \textsc{Power Law + fixed Peak} mass model (blue) and the \textsc{Power Law + evolving Peak} model (green), assuming the mass-dependent spin model in both cases.

The \textsc{Power Law + fixed Peak} model is slightly preferred by a Bayes factor of 1.55.
The two models produce results that are broadly consistent, though, there are noticeable differences. Marginalizing over the redshift evolution of the Gaussian peak, the \textsc{Power Law + evolving Peak} model results in $H_0=45^{+68}_{-24}$. We interpret the shift of $H_0$ towards smaller values as compensation for the non-vanishing support at nonzero $\xi_z$.  
When $H_0$ is small, the GW catalog probes a much smaller range of redshifts, meaning that regardless of $\xi_z$, the product of $\xi_z z$---which determines the fractional change in the peak location with redshift---is always small.
Thus, nonzero $\xi_z$ is indicative of small $H_0$, so that the measurable evolution of the mass distribution remains small. 
Because a smaller $H_0$ shifts the inferred redshifts of the GW events to smaller values, the corresponding merger rate also has more support at lower redshifts. 
The posterior on $\gamma$, which governs the slope of the merger rate in the low-redshift regime, shifts towards smaller values under the \textsc{Power Law + evolving Peak} model.

\subsection{Simulated data results}
To demonstrate the effectiveness of using multiple population features in a spectral siren inference, we perform our analysis on simulated data.
We consider the case in which the Gaussian peak component of the mass distribution evolves with redshift, taking $\xi_{z}=0.1$ for our simulated catalog. As we show in Sec.~\ref{sec:GWTC3-results} and in Appendix~\ref{appendix:evolving peak population result}, this degree of mass evolution with redshift is consistent with GWTC-3 under either loose or strict cosmological assumptions. We assume a catalog of 300 BBH events with O3b-like selection effects. Rather than performing full parameter estimation on the simulated events, for our simple demonstrative study, we assume no measurement uncertainty. Additional details regarding the injected population are described in Appendix~\ref{appendix:injection model}.
\begin{figure}
    \centering
    \includegraphics[width=1\linewidth]{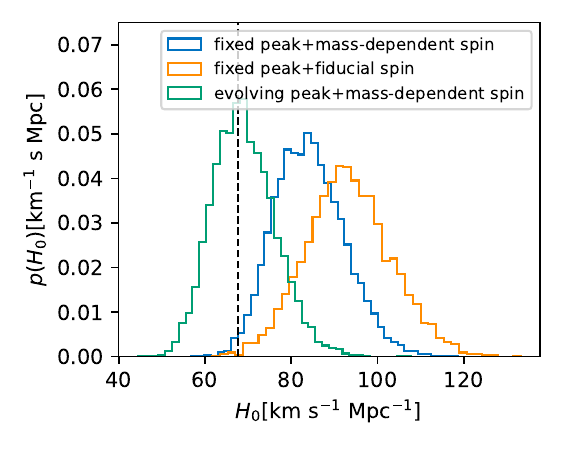}
    \caption{$H_0$ posteriors on simulated data using different mass and spin models. The simulation is generated from \textsc{Power Law + evolving Peak} mass and the mass-dependent spin models. The vertical dashed line is the true value of $H_0$.}
    \label{fig:injection_nomulti_evolve_fixed_comparison}
\end{figure}

In Fig.~\ref{fig:injection_nomulti_evolve_fixed_comparison}, we show the inferred $H_0$ under different population model assumptions, according to the standard inference in which all population properties share an identical $H_0$. As a benchmark, in green we show the correct inference in which we self-consistently use the injected population model in the fit, i.e., the \textsc{Power Law + evolving Peak} mass model and the mass-dependent spin model. As expected, the true value of $H_0=67.7\ \mathrm{ km \ s^{-1} \ Mpc^{-1}}$ is reconstructed well under the correct model assumptions.
In blue, we show the less optimal scenario in which we neglect the redshift evolution of the mass distribution, using the \textsc{Power Law + fixed Peak} mass model, but correctly model the spin-mass correlation, which slightly alleviates some of the bias of the misspecified mass model. 

The worst-case scenario is shown in orange: we neglect both the redshift evolution of the mass distribution and the spin-mass correlation, using the \textsc{Power Law + fixed Peak} mass model and the fiducial spin model. 
In both cases where the redshift evolution of the mass distribution is neglected in the fit, the $H_0$ posterior is biased towards larger values. In the worst case, the true value is ruled out at 99\% credibility, while correctly fitting the mass-spin correlation includes the true value within the 98\% credible interval.

\begin{figure}
    \centering
    \includegraphics[width=1\linewidth]{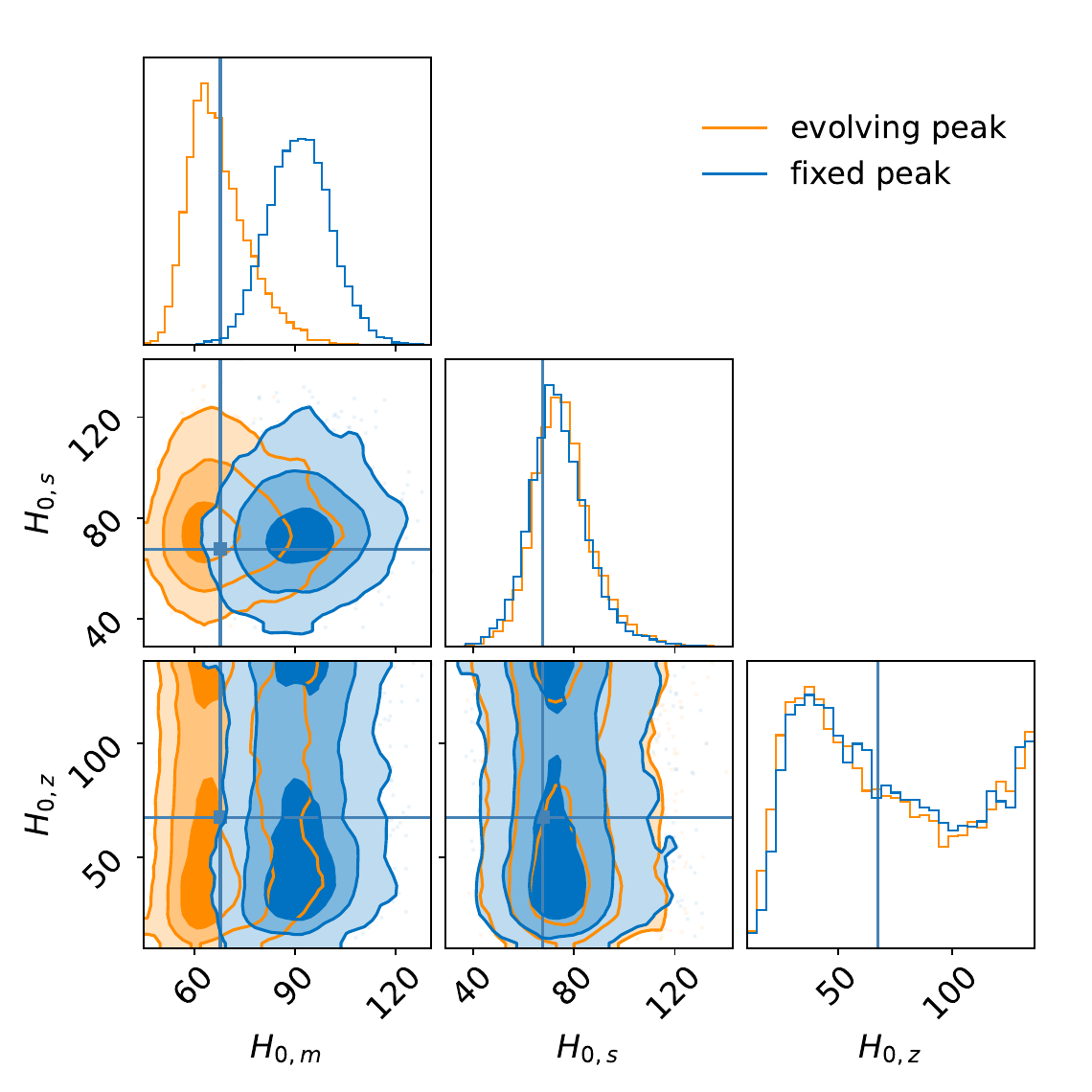}
    \caption{Constraints on $H_0$ from different parts of population properties using \textsc{Power Law + fixed Peak} mass model or \textsc{Power Law + evolving Peak} mass model on simulated data. The mass-dependent spin model is used for both analyses.
    The Hubble parameter $H_0$ is measured in units of $\unit[]{km\,s^{-1}Mpc^{-1}}$.
    }
    \label{fig:injection_69_H0_multi}
\end{figure}

In Fig.~\ref{fig:injection_69_H0_multi}, we demonstrate that our strategy of separating out the contribution to the $H_0$ measurement from different population properties can effectively mitigate against bias from model misspecification. Using the same simulated data, we fit $H_{0,m}$, $H_{0,s}$ and $H_{0,z}$ under the correct \textsc{Power law + evolving peak} model (orange) and the incorrect \textsc{Power law + fixed peak} model (blue). The incorrect mass model assumption leads to significant bias in $H_{0,m}$ as shown in blue. However, the mass-dependent spin model provides an unbiased, informative $H_{0,s}$ inference, although its contribution to the total posterior in the shared $H_0$ case is not significant, since the inference is dominated by $H_{0, m}$. 

We contend that if one can confirm more trustworthy features in the mass distribution, the method of multiple $H_0$ can isolate information only from those more robust features and avoid systematic bias from potentially wrong assumptions. In this case, the mass-spin correlation provides unbiased cosmological inference on $H_{0,s}=73^{+22}_{-17}$ even using incorrect \textsc{Power Law + fixed Peak} mass model. The result is almost identical to the analysis using the correct mass model, which leads to $H_{0,s}=75^{+23}_{-17}$, shown in orange in Fig.~\ref{fig:injection_69_H0_multi}.
The misspecified mass assumption does not bias the $H_{0,s}$ inference because the spin model in this method depends on the detector-frame mass distribution and its own independent cosmological parameters. The extra flexibility of the free cosmological parameters allows the overall detector-frame mass distribution to fit the data sufficiently well regardless of the mass model assumptions.

\section{Discussion and conclusions}\label{sec:disucssion}
In this paper, we update the GWTC-3 spectral siren results using a mass-dependent spin model, which introduces an additional source-frame mass scale that serves as a cosmological ruler. We show that with current data, the inclusion of this spin-mass correlation does not significantly improve the $H_0$ measurement, consistent with the results of \cite{Li_2024}, who performed a similar spectral siren study with multiple subpopulations in mass and spin.
We also explore how the potential redshift evolution of the BBH mass distribution influences the GWTC-3 spectral siren results. Considering a model that allows  the Gaussian peak in the mass distribution to evolve linearly with redshift, we find that the inferred $H_0$ posterior is centered at a lower value from the fiducial \textsc{Power Law + fixed Peak} result, although they remain consistent with each other within uncertainties. There is no conclusive evidence for the presence or absence of a redshift-mass correlation; the the non-evolving \textsc{Power Law + fixed Peak} model is slightly favored by a Bayes factor of 1.55. A similar analysis was performed in \cite{Karathanasis_2023}, where a more astrophysically motivated model was adopted to study the impact of a redshift-dependent mass model on the GWTC-3 spectral siren result. Their results are consistent with ours, with slight differences due to different modeling approaches and event selection. 

We show in Appendix~\ref{appendix:evolving peak population result} that on the condition of strict cosmological assumptions, a flat $\Lambda$CDM Universe with $H_0=67.7$ and $\Omega_m=0.3065$, the possibility of redshift evolution of the Gaussian peak in mass distribution is not completely ruled out with $\xi_z=0.14^{+0.36}_{-0.32}$, consistent with \cite{lalleman2025}.
More observations are needed to determine whether the mass distribution depends on redshift and to what extent .
Additional model misspecification may be lurking beneath the surface, and the degeneracies between arbitrary mass-redshift correlations and cosmological parameters should be further studied.

We propose the idea of assigning independent cosmological parameters to different features in the spectral siren analysis. This allows us to disentangle the impact of different modeling choices, identify possible systematics, and use only the features we trust. 
We show how different cosmological parameters can be naturally assigned to the mass, spin and redshift distributions of BBH events. There are potential variations of this approach. For example, different sets of cosmological parameters can be assigned to different components of the mass distribution, e.g. peaks or edges. This is particularly useful in the case when the regular spectral siren analysis may be dominated by untrustworthy features, and we wish to discard these spurious sources of information.
Our method allows us to marginalize over these features and mitigate the impact of model misspecification. Unbiased cosmological constraints can be inferred using only the more robust features in the population.

We demonstrate the effectiveness of this method in our injection study. We show that by making an incorrect assumption and neglecting the redshift evolution of the mass distribution, the default spectral siren analysis in which all parameters share the same $H_0$ is significantly biased by the strong features in mass distribution, similar to the results in \cite{Ezquiaga_2022,Pierra_2023}. However, we can isolate the cosmological information from only the spin-mass correlation if we assign different $H_0$ to different features in the population, which helps us get unbiased cosmological inference thanks to the extra flexibility of our model.
Moreover, our method can be utilised to check for the consistency of cosmological constraints between different spectral siren features. Inconsistency of the cosmological parameters inferred from different properties would flag potential issues like model misspecification.

The transition mass in the BBH spin distribution would be relatively redshift-invariant in the source frame if it is indeed the lower edge of the PISN mass gap.
We therefore argue that it is a robust feature for spectral siren cosmology. In general, because BBH spins are not redshifted in the detector frame and their correlation with mass likely stems from source-frame physics, features in the joint mass-spin distribution are promising for spectral siren cosmology, although caution is warranted by potentially falsified mass-spin correlations from other features, such as redshift-spin correlations \citep{Biscoveanu_2022}.
With the rapidly growing GW catalog, we look forward to fully utilizing the joint mass-spin-redshift BBH distribution for spectral siren cosmology, where multiple population features can be used not only to improve the constraints, but also to cross-check each other and mitigate systematic bias. 

\begin{acknowledgments}
We thank Adith Praveen for helpful discussions.
This work is supported through Australian Research Council (ARC) Centres of Excellence CE170100004, CE230100016, Discovery Projects DP220101610 and DP230103088, and LIEF Project LE210100002.
MF acknowledges support from the Natural Sciences and Engineering Research Council of Canada (NSERC) under grant RGPIN-2023-05511, the University of Toronto Connaught Fund, and the Alfred P. Sloan Foundation. 
This material is based upon work supported by NSF's LIGO Laboratory which is a major facility fully funded by the National Science Foundation.
The authors are grateful for for computational resources provided by the LIGO Laboratory computing cluster at California Institute of Technology supported by National Science Foundation Grants PHY-0757058 and PHY-0823459, and the Ngarrgu Tindebeek / OzSTAR Australian national facility at Swinburne University of Technology.
LIGO was constructed by the California Institute of Technology and Massachusetts Institute of Technology with funding from the National Science Foundation and operates under cooperative agreement PHY-1764464. This paper carries LIGO Document Number LIGO-P2400105.
\end{acknowledgments}

\appendix

\section{Alternative dataset}\label{appendix:alternative dataset}
In this Appendix, we show the spectral siren results of datasets chosen by different thresholds from \cite{gwtc-3_pop} and \cite{gwtc3_cosmo}. We consider detections by a hard cutoff of minimum mass of components mass to be $>3M_\odot$, i.e., classified as a BBH merger, under LVK default cosmology model. There are 42 BBHs with network matched filter signal-to-noise ratio $>$ 11 and Inverse False Alarm Rate (IFAR) $>\unit[4]{yr}$ selected by \cite{gwtc3_cosmo} while there are 69 BBHs with $\text{IFAR}>\unit[1]{yr}$ selected by \cite{gwtc-3_pop}.

\begin{figure}
    \centering
    \includegraphics[width=1\linewidth]{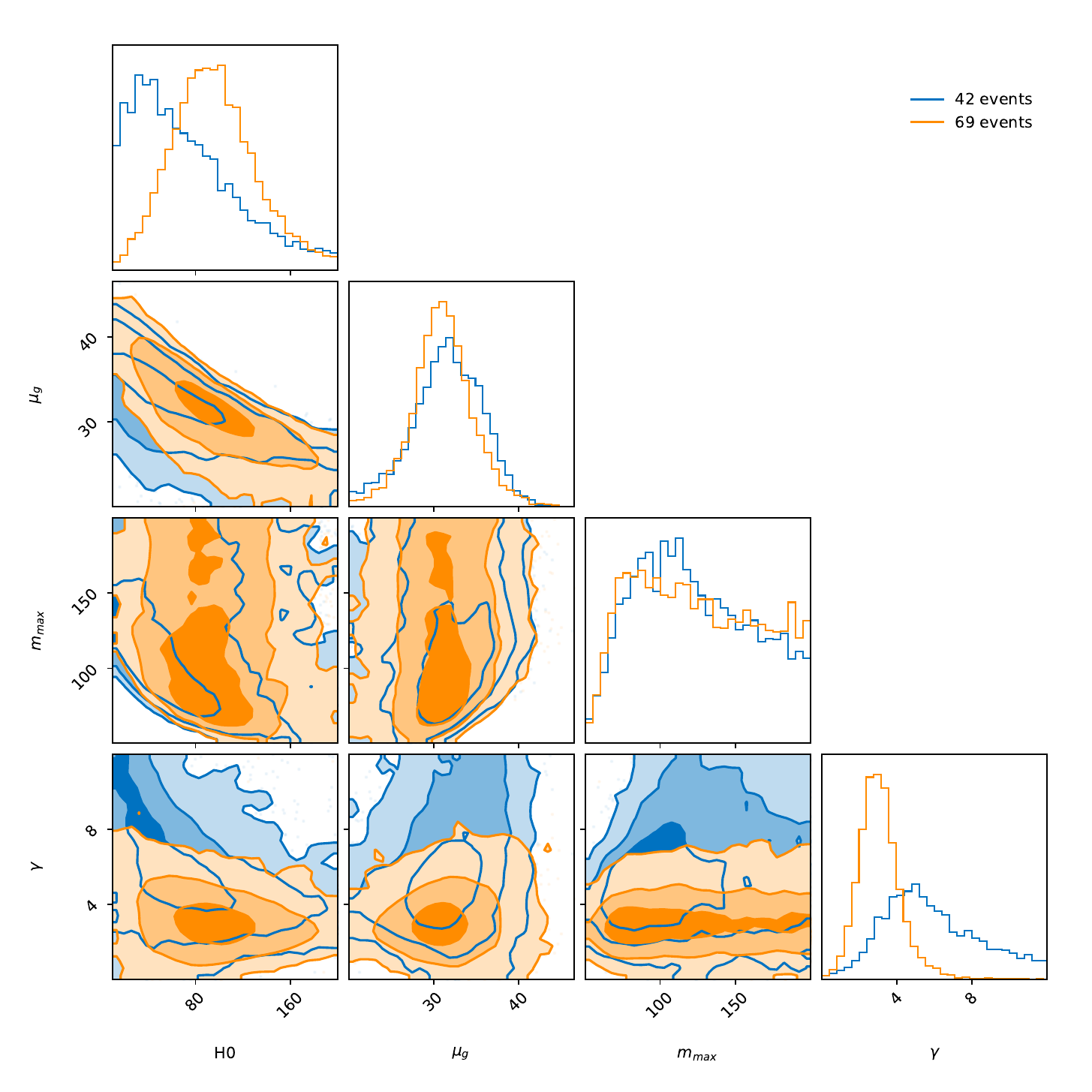}
    \caption{Posterior for $H_0$ and population parameters $\mu_g$, $m_{\rm{max}}$ and $\gamma$ using datasets selected from different thresholds.}
    \label{fig:42_69_comparison}
\end{figure}

To be consistent with the analyses in \cite{gwtc3_cosmo}, we consider \textsc{Power Law + fixed Peak} model (\textsc{Power Law + Peak} model in \cite{gwtc3_cosmo}), fiducial spin model and \cite{Madau_2014} like redshift model. Fig.~\ref{fig:42_69_comparison} shows the posterior distribution between the cosmological parameters and the parameters $\mu_g$ and $m_{\rm{max}}$ as defined in Eq. (A11) in \cite{gwtc3_cosmo}, which govern the location of the BBH Gaussian excess and the upper end of the mass distribution, and the low-redshift power law slope $\gamma$.

The results are consistent with each other. The inferred $H_0$ changes from $66^{+91}_{-46}$ to $93^{+67}_{-52}$ when more detections are included by a less strict threshold. The low-redshift power law slope also shows some difference as higher merger rate in low redshift is more favored in analysis using 69 events.

\section{GWTC-3 \textsc{Power Law + evolving Peak} model population analysis result}\label{appendix:evolving peak population result}
\begin{figure}
    \centering
    \includegraphics[width=1\linewidth]{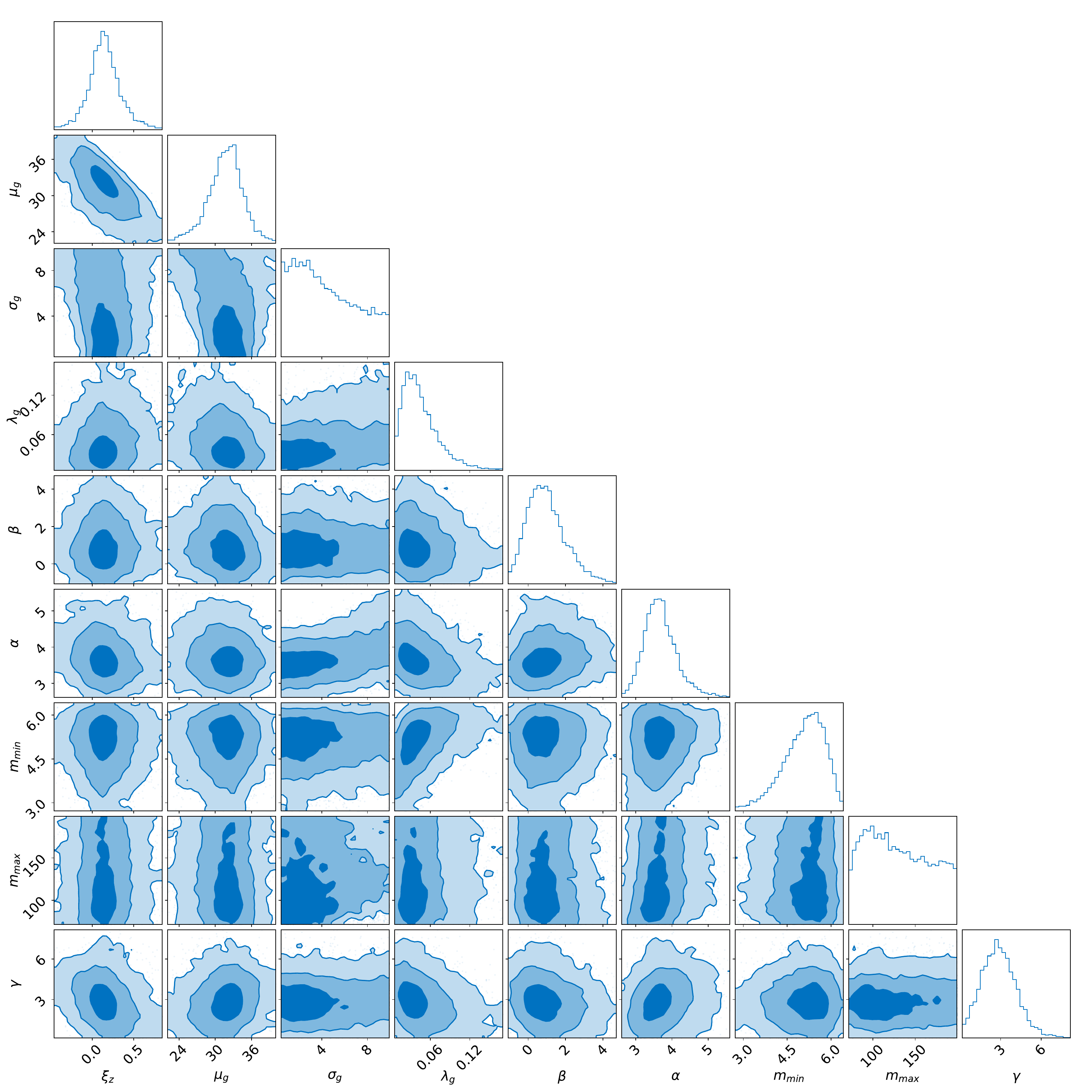}
    \caption{Posterior for main population parameters of \textsc{Power Law + evolving Peak} model and the low-redshift power law slope in the analysis using GWTC-3 data when cosmological parameters are fixed.}
    \label{fig:population_xi}
\end{figure}

We show GWTC-3 \textsc{Power Law + evolving Peak} model population analysis result in this Appendix considering a fixed cosmological assumption. We consider a flat $\Lambda CDM$ Universe with free $H_0=67.7$ and $\Omega_m=0.3065$. We use the mass-dependent spin model and \cite{Madau_2014} like redshift model.
In Fig.~\ref{fig:population_xi}, we show the posterior distribution of the main population parameters.

\section{Injected population model}\label{appendix:injection model}
We describe the injected population properties of our simulated catalog here.
We simulate the masses of the events following \textsc{Power Law + evolving Peak} model. 
The spin distribution and redshift distribution follows the mass-dependent spin model and \cite{Madau_2014} like redshift model described in Eq.~\ref{model:spin} and Eq.~\ref{model:redshift}. We assume a flat $\Lambda$CDM Universe with $\Omega_m=0.3065$ and $H_0=67.7$. The true values of other population parameters are as follows $\alpha=3.74, \beta=1.34,\lambda_g=0.03,\xi_z=0.1,\mu_g=34,\sigma_g=2.5,m_{\rm{min}}=5.17,m_{\rm{max}}=97.91,\delta_m=5.4,\mu=0.03,log\sigma=-1.1,\mu_u=0.55,log\sigma_u=-1.65,w=0.6,m_t=44,\gamma=2.69,\kappa=3.52,z_p=4$.

\section{Prior assumptions of hyper-parameters}\label{appendix:priors}
In Table.~\ref{tab:priors}, we describe the priors of hyper-parameters in \textsc{Power Law + fixed Peak} or \textsc{Power Law + evolving Peak} model, mass-depedent spin model and \cite{Madau_2014} like redshift model used in the analyses in this paper, following \cite{gwtc3_cosmo} and \cite{antonini2024}.

\begin{table}[h!]\label{tab:priors}
    \centering
    \begin{tabular}{llll}
        \hline
        {} & \multicolumn{3}{c}{\textsc{Power Law + Fixed (Evolving) Peak Model}} \\
        \hline
        {\bf Parameter} & \textbf{Description} & & \textbf{Prior} \\
        \hline \hline
        $\alpha$ & Spectral index for the PL of the primary mass distribution. & & $\mathcal{U}(1.5, 12.0)$ \\
        $\beta$ & Spectral index for the PL of the mass ratio distribution. & & $\mathcal{U}(-4.0, 12.0)$ \\
        $m_{\text{min}}$ & Minimum mass of the PL component of the primary mass distribution. & & $\mathcal{U}(2.0\ \text{M}_\odot, 10.0\ \text{M}_\odot)$ \\
        $m_{\text{max}}$ & Maximum mass of the PL component of the primary mass distribution. & & $\mathcal{U}(50.0\ \text{M}_\odot, 200.0\ \text{M}_\odot)$ \\
        $\lambda_{\text{g}}$ & Fraction of the model in the Gaussian component. & & $\mathcal{U}(0.0, 1.0)$ \\
        $\mu_{\text{g}}$ & Fiducial mean of the Gaussian component in the primary mass distribution. & & $\mathcal{U}(20.0\ \text{M}_\odot, 50.0\ \text{M}_\odot)$ \\
        $\sigma_{\text{g}}$ & Width of the Gaussian component in the primary mass distribution. & & $\mathcal{U}(0.4\ \text{M}_\odot, 10.0\ \text{M}_\odot)$ \\
        $\xi_{z}$ & The slope of redshift evolution of the peak & & $\mathcal{U}(0, 2)$ \\
        $\delta_{m}$ & Range of mass tapering at the lower end of the mass distribution. & & $\mathcal{U}(0.0\ \text{M}_\odot, 10.0\ \text{M}_\odot)$ \\
        \hline
        {} & \multicolumn{3}{c}{\textsc{Mass-dependent Spin Model}} \\
        \hline
        {\bf Parameter} & \textbf{Description} & & \textbf{Prior} \\
        \hline \hline
        $m_t$ & Transition mass of different spin distribution & & $\mathcal{U}(20\ \text{M}_\odot, 100\ \text{M}_\odot)$ \\
        $w$ & Width of uniform component of spin distribution for mergers above transition mass & & $\mathcal{U}(0, 1)$ \\
        $\mu$ & Mean of the Gaussian distribution of spin for mergers below transition mass & & $\mathcal{U}(-1, 1)$ \\
        $\xi$ & Fraction of the uniform component of spin distribution for mergers above transition mass & & $\mathcal{U}(0, 1)$ \\
        $\mu_{\text{u}}$ & Mean of the Gaussian distribution of spin for mergers above transition mass & & $\mathcal{U}(-1, 1)$ \\
        $\log_{10}\sigma$ & The width of the Gaussian distribution of spin for mergers below transition mass & & $\mathcal{U}(-2, 1)$ \\
        $\log_{10}\sigma_{\text{u}}$ & The width of the Gaussian distribution of spin for mergers above transition mass & & $\mathcal{U}(-2, 1)$ \\
        \hline
        {} & \multicolumn{3}{c}{\textsc{\cite{Madau_2014} Like Redshift Model}} \\
        \hline
        {\bf Parameter} & \textbf{Description} & & \textbf{Prior} \\
        \hline \hline
        $\gamma$ & Slope of the power-law regime for the rate evolution before the point $z_p$ & & $\mathcal{U}(0, 12)$ \\
        $k$ & Slope of the power-law regime for the rate evolution after the point $z_{\text{p}}$ & & $\mathcal{U}(0, 6)$ \\
        $z_p$ & Redshift turning point between the power-law regimes with $\gamma$ and $k$ & & $\mathcal{U}(0, 4)$ \\
        \hline 
    \end{tabular}
\caption{Priors used for population parameters in each model.}
\end{table}

\bibliography{refs}{}
\bibliographystyle{aasjournal}

\end{document}